\documentclass[aps,prb,twocolumn,superscriptaddress]{revtex4}
\usepackage{graphicx}
\usepackage{bm}
\usepackage{color}

\begin{document}

\preprint{}
%

\title{High-$T_c$ Iron-phosphide Superconductivity Enhanced by Reemergent Antiferromagnetic Spin Fluctuations in (Sr$_4$Sc$_2$O$_6$)Fe$_2$(As$_{1-x}$P$_{x}$)$_2$ probed by NMR }

\author{F.  Sakano}
\author{K. Nakamura}
\author{T.  Kouchi}
\author{T.  Shiota}
\author{F.  Engetsu}
\author{K. Suzuki}
\author{R. Horikawa}
\author{M. Yashima}
\affiliation{Graduate School of Engineering Science, Osaka University, Osaka 560-8531, Japan}
\author{S. Miyasaka}
\author{S. Tajima}
\affiliation{Graduate school of Science, Osaka University, Osaka 560-0043, Japan}
\author{A. Iyo}
\affiliation{National Institute of Advanced Industrial Science and Technology (AIST), Umezono, Tsukuba 305-8568, Japan}
\author{Y. -F. Guo}
\affiliation{Advanced Materials Laboratory, National Institute for Materials Science (NIMS),  Tsukuba, Ibaraki 305-0044, Japan}
\affiliation{School of Physical Science and Technology, ShanghaiTech University, Shanghai 201210, China}
\author{K. Yamaura}
\author{E. Takayama-Muromachi}
\affiliation{Advanced Materials Laboratory, National Institute for Materials Science (NIMS),  Tsukuba, Ibaraki 305-0044, Japan}
\author{M. Yogi}
\affiliation{Faculty of Science, University of the Ryukyus,  Okinawa 903-0213, Japan}
\author{H. Mukuda}\email[]{e-mail  address: mukuda@mp.es.osaka-u.ac.jp}
\affiliation{Graduate School of Engineering Science, Osaka University, Osaka 560-8531, Japan}

\date{\today}

\begin{abstract}

We report a systematic NMR study on [Sr$_4$Sc$_2$O$_6$]Fe$_2$(As$_{1-x}$P$_x$)$_2$, for which the local lattice parameters of the iron-pnictogen (Fe$Pn$) layer are similar to those of the series LaFe(As$_{1-x'}$P$_{x'}$)O, which exhibit two segregated antiferromagnetic (AFM) order phases, AFM1 at $x'$=0-0.2 and AFM2 at $x'$=0.4-0.7.  
Our results revealed that the parent AFM1 phase at $x$=0 disappears at $x$=0.3-0.4, corresponding to a pnictogen height ($h_{pn}$) from the Fe-plane of 1.3-1.32 \AA, which is similar to that of LaFe(As$_{1-x'}$P$_{x'}$)O and various parent Fe-pnictides. 
By contrast, the AFM2 order reported for LaFe(As$_{0.4}$P$_{0.6}$)O does not appear at $x\sim$0.8, although the local lattice parameters of the Fe$Pn$ layer and the microscopic electronic states are quite similar.
Despite the absence of the {\it static} AFM2 order, reemergent {\it dynamical} AFM spin fluctuations were observed at approximately $x\sim$0.8, which can be attributed to the instability of the AFM2 phase. 
We suggest this re-enhancement of AFM spin fluctuations to play a significant role in enhancing the $T_c$ to 17 K for $x$=0.8-1.
Finally, we discuss the universality and diversity of the complicated magnetic ground states from a microscopic point of view, including the difference in the origins of the AFM1 and AFM2 phases, and their relations with the high superconducting transitions in Fe-pnictides.

\end{abstract}


\maketitle

\section{Introduction}

Since the discovery of high-temperature superconductivity (SC) in iron(Fe)-based compounds\cite{Kamihara2008}, a number of researches have unraveled a rich variety of antiferromagnetic (AFM), structural, and superconducting phase diagrams of various Fe-pnictide($Pn$)/chalcogen families. 
These phase diagrams are drastically changed by the local lattice parameters and carrier density of the Fe$Pn$ layers\cite{IshidaRev,Stewart,Scalapino,Hosono_review}. 
Optimization of both the local lattice parameters and the electron/hole-doping levels of the Fe$Pn$/chalcogen layer is necessary to raise the SC transition temperature ($T_c$) to above 50 K.\cite{IshidaRev,Stewart,Scalapino,Hosono_review,Ren1,C.H.Lee,Mizuguchi,Wang,He,Miyata,Zhao}
The parent materials of Fe-based superconductors characterized by the formal Fe$^{2+}$ valence state such as in LaFeAsO, exhibit an AFM order in association with the orthorhombic transition, which is denoted as AFM1 hereafter. 
The isovalent substitution of P for As causes the local lattice parameters of the Fe$Pn$ layers to undergo deformation without variation of the Fe$^{2+}$ state.
The pnictogen height ($h_{Pn}$) from the Fe-plane provides one possible classification of the ground states for the parent and its isovalent-substituted Fe-pnictides (See Fig. \ref{h_pn}). This indicates that the AFM1 phase prevails when 1.32 \AA $<h_{Pn}<$1.42 \AA\, which separates the nodeless SC state at $h_{Pn}>$1.42 \AA\ and the nodal SC state at $h_{Pn}<$1.3 \AA. The separation seems to be insensitive to the Fe-Fe bonding length ($d_{\rm Fe-Fe}$)\cite{KinouchiPRB}.

Recently, the reemergent AFM order phase separated from AFM1 of LaFeAsO was observed in the range $0.4\le x'\le 0.7$ of LaFe(As$_{1-x'}$P$_{x'}$)O\cite{SKitagawa_2014,Lai_PRB,Miyasaka,Mukuda_jpsj2014}, and is denoted as AFM2 hereafter. 
Furthermore, another type of segregated AFM order phase was also reported in heavily electron-doped $Ln$FeAs(O$_{1-y}$H$_{y}$) for $y\sim0.5$\cite{Iimura_H,Hiraishi,IimuraSm}, and is denoted as AFM3 here.
To unravel the universality/diversity of the emergent phases, here we focus on the series [Sr$_4$Sc$_2$O$_6$]Fe$_2$(As$_{1-x}$P$_{x}$)$_2$, for which the local lattice parameters of the Fe$Pn$ layer are similar to those of the series LaFe(As$_{1-x'}$P$_{x'}$)O.
The series [Sr$_4$Sc$_2$O$_6$]Fe$_2$(As$_{1-x}$P$_x$)$_2$, denoted as SrSc42622(As$_{1-x}$P$_x$) hereafter, was previously reported to show the AFM order below $T_{\rm N}$=35 K for the compound with $x$=0 \cite{Munevar}, whereas the compound with $x$=1 is a superconductor with a possible nodal gap below the {\it onset} $T_c^{onset}\sim$17 K\cite{Ogino,Yates}. However,   
an investigation of the intermediate region between $x$=0 and 1 has not been reported thus far. 
It is noteworthy that the $T_c$ for $x$=1 of SrSc42622(As$_{1-x}$P$_x$) is remarkably high among the various iron-phosphide (FeP) end members, e.g., LaFePO ($T_c$= 6 K)\cite{Kamihara2006}, LiFeP($T_c$= 5 K)\cite{Deng}, [Ca$_4$Al$_2$O$_6$]Fe$_2$P$_2$(=CaAl42622(P))($T_c$= 17 K)\cite{Shirage_AsP}, and $Ae$Fe$_2$P$_2$($Ae$=Ba,Sr,Ca)(non-SC).
Further systematic studies over a wide range of $x$ in SrSc42622(As$_{1-x}$P$_x$)  provide an opportunity to unravel the origin of the high $T_c$ state, the universality of their ground states, and the relationship between local lattice parameters and some segregated AFM and SC phases.

In this paper, we report systematic $^{75}$As and $^{31}$P-NMR studies of SrSc42622(As$_{1-x}$P$_x$) for 0 $\le x\le$ 1 and compare the outcome with previous results on the various parent and isovalent-substituted Fe-pnictides.
As a result, we reveal that (i) the AFM1 phase in parent Fe-pnictides disappears  when  $h_{pn}\le$1.3-1.32 \AA, which is insensitive to  $d_{\rm Fe-Fe}$, (ii) the {\it static} AFM2 phase reported for  LaFe(As$_{0.4}$P$_{0.6}$)O does not appear in the series [Sr$_4$Sc$_2$O$_6$]Fe$_2$(As$_{1-x}$P$_x$)$_2$  despite the similarity of the local lattice parameters of the Fe$Pn$ layer.
Instead, we revealed that the re-enhanced AFM spin fluctuations were derived from the possible instability of the AFM2 order, which plays a significant role in achieving the highest-$T_c$ state ($T_c$=17 K) at $x=$0.8$\sim$1 among the phosphorous-rich Fe-based superconductors. 
We discuss the universality and diversity of their complicated ground states and the high SC transitions in Fe-pnictides from a microscopic point of view.

\section{Experimental }

NMR measurements were performed on coarse-powder polycrystalline samples of [Sr$_4$Sc$_2$O$_6$]Fe$_2$(As$_{1-x}$P$_x$)$_2$ with nominal contents for $x$=0, 0.2, 0.4, 0.6, 0.8, and 1.0. The samples were synthesized by a solid-state reaction method~\cite{Zhang}. 
Bulk $T_c$s were determined from the onset of SC diamagnetism in the susceptibility measurement, which revealed $T_c\sim$17 K for $x$=0.8 and $T_c\sim$13 K\cite{Tconset} for $x$=1.0. 
No SC transition was observed in the range $0\le x \le0.6$ from the susceptibility measurement.
The parent compound ($x$=0) was investigated by 
$^{75}$As (nuclear spin $I$=3/2) and $^{45}$Sc-NMR ($I$=7/2).
For $x\ge 0.2$, the Knight shift ($K$) and the nuclear-spin lattice-relaxation rate $^{31}(1/T_1)$ in the normal state was measured by $^{31}$P-NMR ($I$=1/2) mainly at a high magnetic field of $\sim$11.93 T. 
Here $K$ was calibrated using the resonance field of $^{31}$P in H$_3$PO$_4$, and $^{31}(1/T_1)$  was determined by fitting the recovery curve for $^{31}$P nuclear magnetization to a single exponential function $\propto \exp \left(-t/T_1\right)$. 
The SC states of $x$=0.8 and 1.0 were also investigated by means of the $^{31}$P-NMR Knight shift and $^{31}(1/T_1)$ at lower field of $\sim$1 T, which is lower than the upper critical field $B_{c2}$.  
The bulk SC transition in these compounds are corroborated not only by decreases in $K$ but also by increases in the line widths of the spectra below $T_c$(1T) at 1 T.
The values of $h_{Pn}$ for the intermediate $x$ region of SrSc42622(As$_{1-x}$P$_x$) were assumed by interpolation of the data at $x$=0 and 1.0.\cite{Ogino,Munevar,Zhang}

\begin{figure}[htbp]
\centering
\includegraphics[width=8cm]{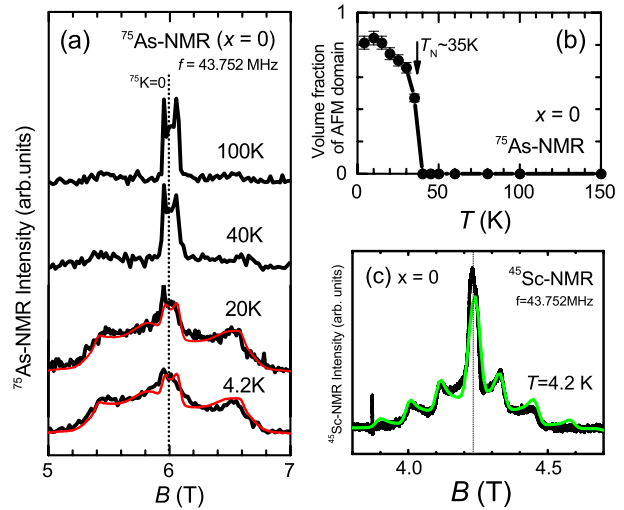}
\caption[]{(Color online)  
(a) $T$ dependence of $^{75}$As-NMR spectra of $x$=0. 
The solid curves for the spectra below 35 K are the results of simulation for an internal field of $^{75}B_{\rm int}$=$\pm$0.5T and  $^{75}\nu_Q$=8.8 MHz at the As site, taking the presence of subtle ingredients of the paramagnetic domains into account.  
(b) $T$ dependence of the volume fraction of AFM domains evaluated from the fractional intensity of the broad spectra in (a), which increases significantly below $T_{\rm N}$=35 K. 
(c) $^{45}$Sc-NMR spectra at 4.2 K  reproduced by simulation (solid curve) for the case of zero internal field at the $^{45}$Sc site in the blocking layer.
}
\label{AsScNMR}
\end{figure}

\section{Results and Discussion}

\subsection{Sr$_4$Sc$_{2}$O$_{6}$Fe$_2$As$_2$ ($x$=0)} 
\subsubsection{AFM order  probed by $^{75}$As and $^{45}$Sc-NMR} 

Figure~\ref{AsScNMR}(a) shows the temperature($T$) dependence of the $^{75}$As-NMR ($I$=3/2) spectra of the powder sample of $x$=0. 
Generally, the Hamiltonian for a nuclear spin with $I$ ($\ge$1) is described by the Zeeman interaction due to the magnetic field $B$ (${\cal H}_{\rm Z}$) and the nuclear-quadrupole interaction (${\cal H}_{\rm Q}$) as follows:
\begin{equation}
{\cal H}={\cal H_{\rm Z}}+{\cal H_{\rm Q}}=-\gamma_{\rm N}\hbar {\bm I} \cdot {\bm B}+\frac{e^{2}qQ}{4I(2I-1)}(3I_{\rm z^{\prime}}^2-I(I+1)),
\label{eq:hamiltonian}
\end{equation}
where $\gamma_{\rm N}$ is the nuclear gyromagnetic ratio, $eQ$ is the nuclear quadrupole moment, and $eq$ is the electric field gradient (EFG) at the nuclear site. 
Here, the nuclear quadrupole resonance (NQR) frequency is defined as $\nu_{\rm Q}=3e^{2}qQ/2h I(2I-1)$, and the asymmetric parameter ($\eta$) is zero for the tetragonal symmetry. 
As shown in Fig. \ref{AsScNMR}(a), above 35 K, the spectrum shows a typical powder pattern for a paramagnetic state, in which the spectral shape affected by the nuclear quadrupole interaction enables us to evaluate the NQR frequency of the $^{75}$As site ($^{75}\nu_{\rm Q}$) to be $\sim$8.8 MHz. 
The broad spectra below 35 K are due to the onset of the AFM order, because magnetically ordered moments induce the internal magnetic field at nuclear sites. 
It enables us to evaluate the N\'eel temperature as being $T_{\rm N}$=35 K, which coincides with the value reported previously as probed by $\mu$SR and  M\"ossbauer experiments.\cite{Munevar}
The small peak at approximately $^{75}K$=0 below 35 K indicates the presence of subtle ingredients of paramagnetic domains. 
As shown by the solid curves in Fig.~\ref{AsScNMR}(a), the observed spectra are well reproduced by assuming the superposition of the predominant broad spectra of the AFM domains\cite{Yamamoto} with $^{75}B_{\rm int}^{\parallel c}\sim\pm0.5$ T and $^{75}\nu_{\rm Q}\sim$8.8 MHz and the paramagnetic domain with the same $^{75}\nu_{\rm Q}$. 
The volume fraction of the AFM domains evaluated in the simulation develops predominantly below $T_{\rm N}$=35 K, as shown in Fig. \ref{AsScNMR}(b), which is also consistent with the previous report\cite{Munevar}.

Figure \ref{AsScNMR}(c) shows the $^{45}$Sc-NMR ($I$=7/2) spectrum at 4.2 K well below $T_{\rm N}$, which is well articulated in contrast to the broad features of the $^{75}$As-NMR spectrum at the same temperature. 
The solid curve represents  the simulated $^{45}$Sc-NMR spectrum for  $^{45}$Sc-NQR frequency $^{45}\nu_{\rm Q}\sim$2.4 MHz and no internal field ($^{45}B_{\rm int}$=0) at the Sc site in the blocking layer.
The results indicate that the hyperfine field transferred from the Fe site to the Sc site is negligibly small, and 
 the blocking layer composed of [Sr$_4$Sc$_2$O$_6$] does not affect the electronic properties of the FeAs layer. 
This is different from the case of the superconducting compound [Sr$_4$V$_2$O$_6$]Fe$_2$As$_2$, in which the electronic states are modified by the possible magnetism of the V site\cite{Ok}. 

\begin{figure}[htbp]
\centering
\includegraphics[width=7cm]{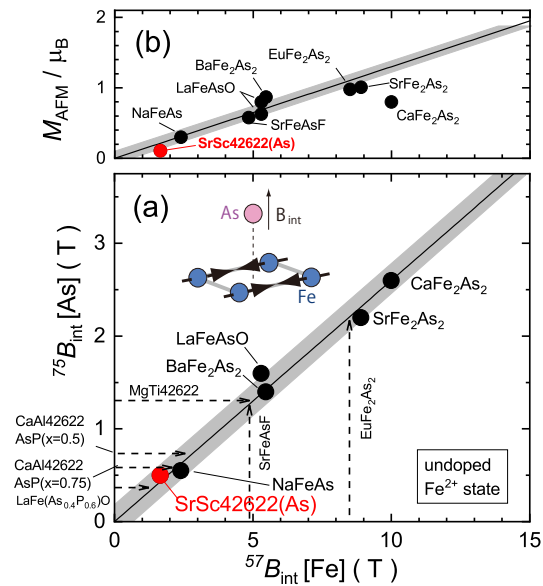}
\caption[]{(Color online)  
(a) Internal field  at the $^{75}$As site ($^{75}B_{\rm int}$) evaluated by NMR and plotted against the internal field at the $^{57}$Fe site ($^{57}B_{\rm int}$) evaluated by $^{57}$Fe M\"ossbauer studies for various undoped Fe-pnictides in the AFM ordered state\cite{IshidaRev,Stewart,Scalapino,Hosono_review,Kitao,H.-F.Li,RotterM,Tegel,Alzamora,FukazawaM,Beak,Tatematsu,Cao,SKitagawa,MukudaFe2,Qureshi,Kitagawa2}
(b)  $^{57}B_{\rm int}$ is proportional to the AFM ordered moment ($M_{\rm AFM}$) estimated by a neutron diffraction experiment.\cite{IshidaRev,Stewart,Scalapino,Hosono_review,Kitao,H.-F.Li,RotterM,Tegel,Alzamora,FukazawaM,Beak,Tatematsu,Cao,SKitagawa,MukudaFe2,Qureshi,Kitagawa2}
The linear relation of these values holds for various Fe pnictides of which the Fe$Pn$ layer possesses different local lattice parameters.
The weak internal field at $x$=0 [SrSc42622(As)] is due to the small $M_{\rm AFM}$.
This relation may enable us to infer the unknown values of $M_{\rm AFM}$, $^{57}B_{\rm int}$, and $^{75}B_{\rm int}$, if one of them is obtained, as shown by the broken arrows in (a). 
}
\label{Hint}
\end{figure}

\subsubsection{ Microscopic evidence of universal behavior in Fe-pnictides  } 

The internal field at the As site ($^{75}B_{\rm int}$) in Fe-pnictides  is mostly induced by an off-diagonal pseudo-dipole field from the stripe-type AFM ordered moment ($M_{\rm AFM}$) lying on the $ab$-plane at the Fe site\cite{Kitagawa1}. 
The value of $^{75}B_{\rm int}$($\sim0.5$ T) for $x$=0 is relatively small among the Fe-pnictides owing to the small $M_{\rm AFM}(\sim0.11\mu_{\rm B}$) evaluated by a neutron diffraction experiment \cite{Munevar}. 
Figure \ref{Hint}(a) shows the $^{75}B_{\rm int}$ derived from $^{75}$As-NMR studies plotted against those at the $^{57}$Fe site ($^{57}B_{\rm int}$) observed by $^{57}$Fe-M\"ossbauer studies for various parent Fe-pnictides. \cite{IshidaRev,Stewart,Scalapino,Hosono_review,Kitao,H.-F.Li,RotterM,Tegel,Alzamora,FukazawaM,Beak,Tatematsu,Cao,SKitagawa,MukudaFe2,Qureshi,Kitagawa2}
The datum of $x$=0 [SrSc42622(As)] is seen on the linear relation between $^{75}B_{\rm int}$, $^{57}B_{\rm int}$, and $M_{\rm AFM}$, as shown in Figs. \ref{Hint}(a) and \ref{Hint}(b).
The slope of the linear relation in Fig. \ref{Hint}(b) enables us to estimate the hyperfine-coupling constants  $^{57}A_{\rm hf}\sim$7.7 T/$\mu_B$, using $^{57}B_{\rm int}=^{57}A_{\rm hf}M_{\rm AFM}$.
The slope of the universal linear relation  between  the $^{57}B_{\rm int}$ and $^{75}B_{\rm int}$,$^{75}B_{\rm int}/^{57}B_{\rm int}$=0.26,  gives  $^{75}A_{\rm hf}\sim$2.0 T/$\mu_B$  using $^{75}B_{\rm int}=^{75}A_{\rm hf}M_{\rm AFM}$. 
It is also noteworthy that these linear relations hold for various Fe pnictides that possess many differences in the local lattice parameters of the Fe$Pn$ layers, such as the Fe-Fe bond lengths, pnictogen heights, and orthorhombicity \cite{Hosono_review}. 
Thus, this relation will help us to deduce the $M_{\rm AFM}$ from the  internal fields  either at   $^{57}$Fe or $^{75}$As sites, even for the case of  lack of the neutron diffraction study. 
For example, as shown by the broken arrows in Fig. \ref{Hint}(a), the AFM moment $M_{\rm AFM}$ at the Fe site is tentatively deduced to be $\sim0.18\mu_{\rm B}$ for LaFe(As$_{0.4}$P$_{0.6}$)O by using the ratio $^{75}\!A_{\rm hf}/^{31}\!A_{\rm hf}$= $^{75}\!B_{\rm int}$/$^{31}\!B_{\rm int}$=3.05 evaluated in [Ca$_4$Al$_2$O$_6$]Fe$_2$(As,P)$_2$~\cite{KinouchiPRB}. 

\begin{figure}[htbp]
\centering
\includegraphics[width=8cm]{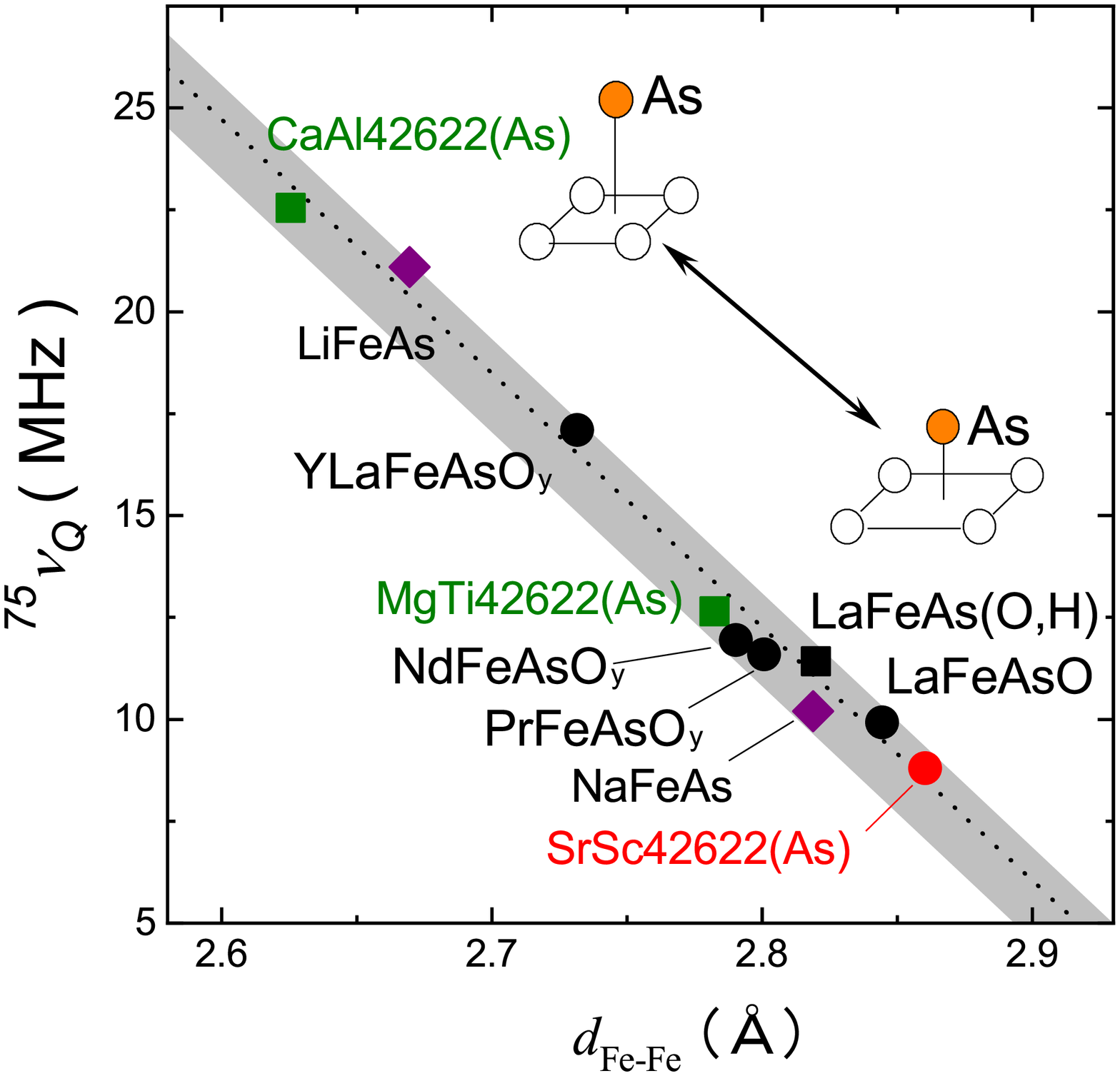}
\caption[]{(Color online)  
 $^{75}$As-NQR frequencies {\it vs.} Fe-Fe bond length $d_{\rm Fe-Fe}$ for $x$=0 [SrSc42622(As)], together with data for $Ln$FeAsO \cite{MukudaNQR,Yamashita_Y,MukudaPRL},  $M$42622\cite{Kinouchi,Yamamoto,KinouchiPRB}, and $A$FeAs($A$=Li,Na) \cite{Li,Kitagawa_Na111}. 
The smallest value of $^{75}\nu_{\rm Q}$ for $x$=0 [SrSc42622(As)] corresponds to the longest $d_{\rm Fe-Fe}$ and lowest $h_{Pn}$ among them.
}
\label{nuQ}
\end{figure}

Further microscopic evidence of the universality between $x$=0 [SrSc42622(As)] and many parent FeAs compounds  can be seen in the relation between $^{75}$As-NQR frequencies ($^{75}\nu_{\rm Q}$) and the  local lattice parameters of the FeAs layer\cite{MukudaNQR,Yamashita_Y,MukudaPRL}. 
As shown in Fig. \ref{nuQ}, the value of  $^{75}\nu_{\rm Q}$ for $x$=0 [SrSc42622(As)] is also linearly related with $d_{\rm Fe-Fe}$($\propto a$-axis length), along with those for $Ln$FeAsO$_{1-y}$\cite{MukudaNQR,Yamashita_Y,MukudaPRL} and  ($Ae_4M_2$O$_6$)Fe$_2$As$_2$ ($M$42622)\cite{Kinouchi,Yamamoto,KinouchiPRB}, and $A$FeAs($A$=Li,Na) \cite{Li,Kitagawa_Na111}, except for the $Ae$Fe$_2$As$_2$-based compounds. 
The value of $^{75}\nu_{\rm Q}$ increases linearly as $d_{\rm Fe-Fe}$($\propto a$-axis length) decreases in many FeAs families, because the value of $^{75}\nu_{\rm Q}$ is proportionally related to the electric field gradient derived from the charge distribution around the $^{75}$As nucleus of the FeAs$_4$ tetrahedron.  
The largest  $^{75}\nu_{\rm Q}$ was observed for CaAl42622(As)\cite{Kinouchi}, which has the shortest $d_{\rm Fe-Fe}$  and highest $h_{Pn}$ among them. 
In contrast, the smallest value of  $^{75}\nu_{\rm Q}$ for $x$=0 [SrSc42622(As)] is additional microscopic evidence that SrSc42622(As) possesses the longest $d_{\rm Fe-Fe}$ and lowest $h_{Pn}$ among them. 
The monotonic variation of $^{75}\nu_{\rm Q}$ for various Fe-pnictides with different local lattice parameters suggests that those of the FeAs layer undergo continuous deformation, which is attributed to the strong covalency of Fe-As bonds that ensures that the Fe-As bond length remains constant\cite{C.H.Lee}.

\begin{figure}[htbp]
\centering
\includegraphics[width=8cm]{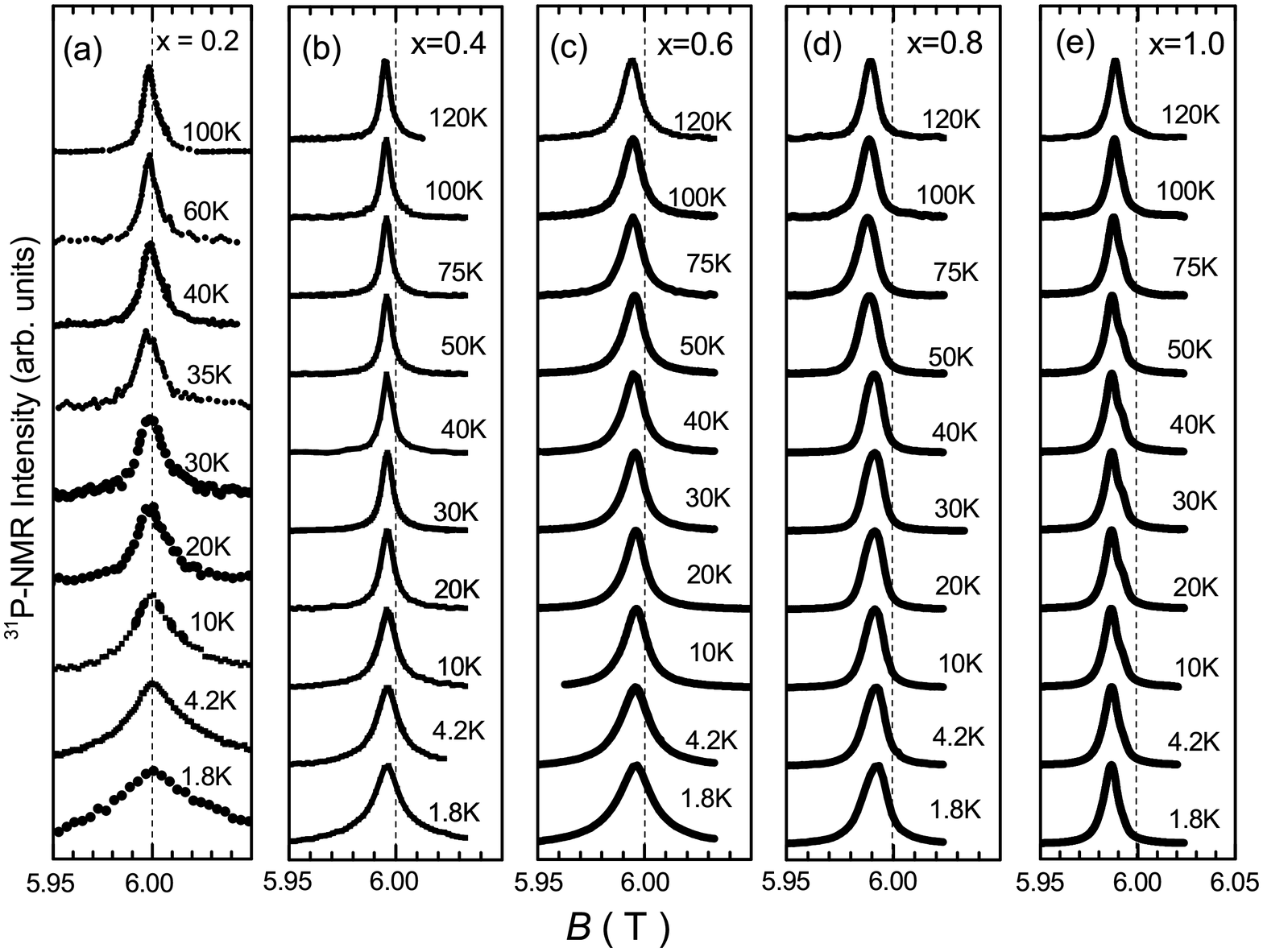}
\includegraphics[width=8.7cm]{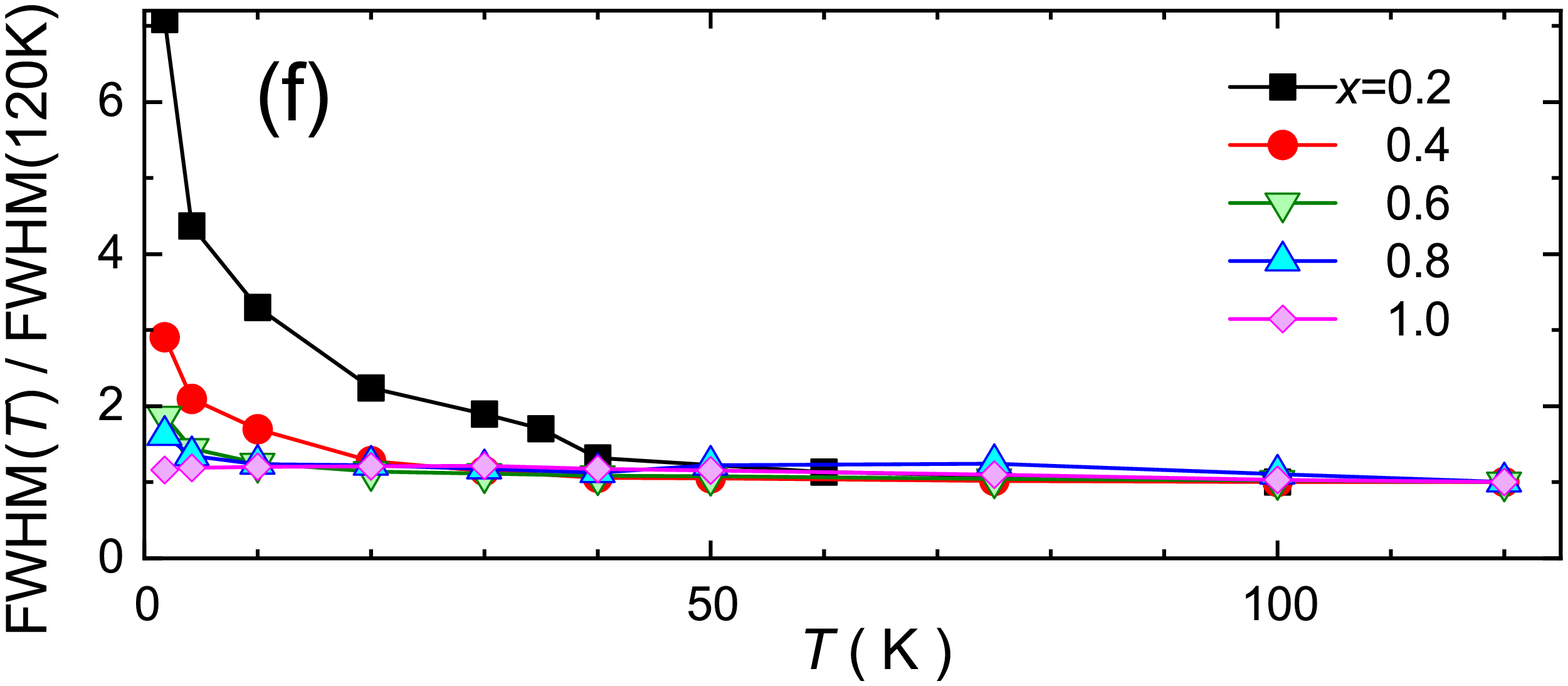}
\includegraphics[width=8cm]{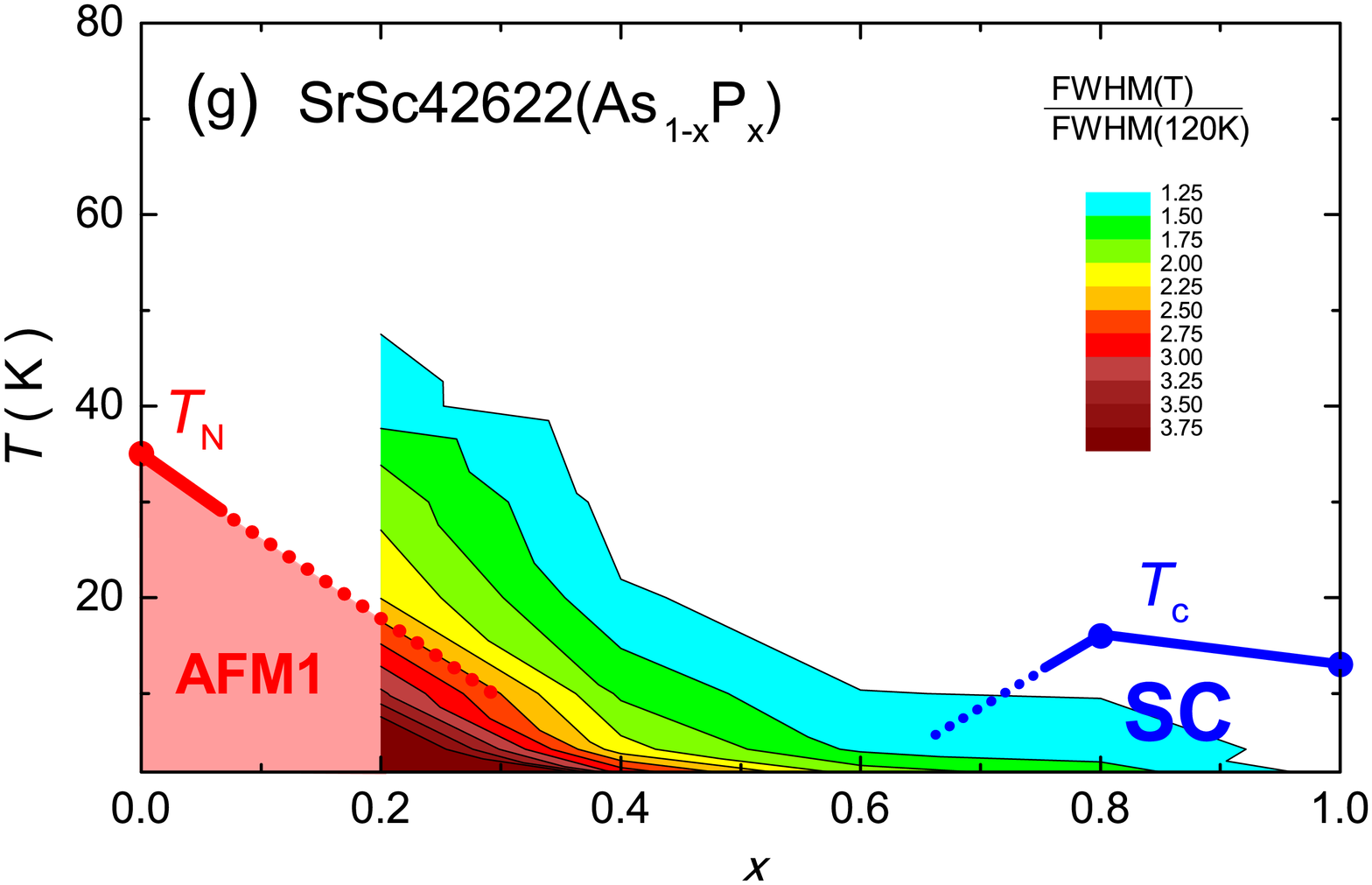}
\caption[]{(Color online)  
 $T$ dependence of field-swept $^{31}$P-NMR spectra for (a) $x$=0.2, (b) 0.4, (c) 0.6, (d) 0.8,  and (e) 1.0  of SrSc42622(As$_{1-x}$P$_{x}$) at a fixed frequency $f_0$=103.422 MHz($B\sim$6 T). 
(f) FWHM at $^{31}$P site normalized by the value at 120 K, which is summarized in (g) the contour plot, revealing the evolution of the internal field at the $^{31}$P site as functions of $T$ and $x$. 
The FWHM increases upon cooling, especially below the broken line extrapolated to $T_{\rm N}$=35 K at $x$=0, suggesting that the AFM1 order phase disappears at $x$=0.3-0.4.  
} 
\label{PNMR}
\end{figure}

\subsection{Sr$_4$Sc$_{2}$O$_{6}$Fe$_2$(As$_{1-x}$P$_{x}$)$_2$ ($x$=0.2-1.0)} 

\subsubsection{Suppression of AFM1 phase probed by $^{31}$P-NMR spectra for 0.2 $\le x\le$1.0 } 

\begin{figure}[tbp]
\centering
\includegraphics[width=8cm]{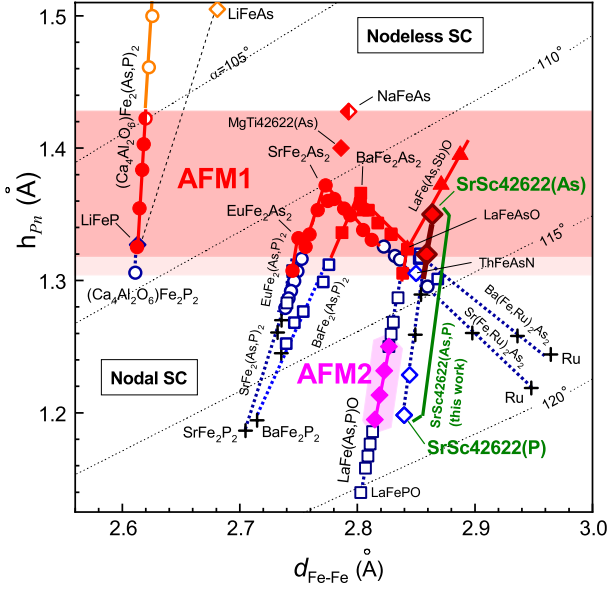}
\caption[]{(Color online)  
Plot of the ground states of parent and isovalent substituted Fe-pnictides with a formal valence of Fe$^{2+}$  as functions of  $h_{Pn}$ and $d_{\rm Fe-Fe}$, which represents the revised situation subsequent to previous reports\cite{KinouchiPRB,Miyamoto}. 
The solid, open, and cross symbols represent the AFM order, the SC, and non-SC states, respectively. 
The data were derived from previous results [Sr$_4$Sc$_2$O$_6$]Fe$_2$(As,P)$_2$\cite{Ogino,Munevar}, LaFe(As,P/Sb)O\cite{C.Wang,Lai_PRB,Mukuda_jpsj2014,Miyasaka,Carlsson}, ThFeAsN\cite{ThFeAsN1,ThFeAsN2,ThFeAsN3}, [Ca$_4$Al$_2$O$_6$]Fe$_2$(As,P)$_2$\cite{Shirage,KinouchiPRB},   [Sr$_4$MgTiO$_6$]Fe$_2$(As,P)$_2$(Mg$_{0.5}$Ti$_{0.5}$42622)\cite{Sato,Yamamoto}, BaFe$_2$(As,P)$_2$\cite{Kasahara}, Ba(Fe,Ru)$_2$As$_2$\cite{Rullier-Albenque,Qiu,Nath}, SrFe$_2$(As,P)$_2$\cite{Kobayashi,KobayashiPRB2013,Miyamoto}, and Sr(Fe,Ru)$_2$As$_2$\cite{Schnelle}, and so on. 
The ground states for 0 $<x<$ 1 of SrSc42622(As$_{1-x}$P$_x$) were obtained in this work.
Note that the {\it static} AFM2 phase, which reemerged in the range $0.4\le x' \le 0.7$ for LaFe(As$_{1-x'}$P$_{x'}$)O, was not observed in $0.6\le x \le 1$ for SrSc42622(As$_{1-x}$P$_x$) even at a comparable value of $h_{pn}$. 
However, the reemergent AFM spin fluctuations are observed in the range $0.6\le x \le 1$(See Fig. \ref{AFM-T1}(b)).
}
\label{h_pn}
\end{figure}

Figures~\ref{PNMR}(a-e) show the $T$ dependence of the $^{31}$P-NMR ($I$=1/2) spectra for 0.2$\le x\le$1.0. 
In the case of (a) $x$=0.2, the $^{31}$P-NMR spectra exhibit significant broadening at low temperatures due to the static internal field at the $^{31}$P site ($^{31}\!B_{\rm int}$) from the Fe site.
Figure  \ref{PNMR}(f) shows the full-width-at-half-maximum (FWHM) at the $^{31}$P site normalized by the value at high temperature ($T$=120 K$\gg$$T_{\rm N}$), which is summarized by the contour plot in Fig. \ref{PNMR}(g). 
These plots reveal that the FWHM values of the $^{31}$P-NMR spectra increase upon cooling, especially below the broken line extrapolated to $T_{\rm N}$=35 K at $x$=0 in the figure.  
These significant increases in FWHM at $x<$0.4 are suppressed for $x\ge$0.4, suggesting that the trace of the parent AFM1 order disappears at approximately $x$=0.3-0.4.  
The $h_{\rm Pn}$ of $x$=0.3-0.4 is approximately $\sim$1.30 \AA, which is similar to the border between the AFM1 and SC phases observed not only for  $x'\sim$0.2 of LaFe(As$_{1-x'}$P$_{x'}$)O\cite{SKitagawa_2014,Mukuda_jpsj2014} but also for many  parent Fe-pnictides with the formal valence of the Fe$^{2+}$ state\cite{KinouchiPRB,Miyamoto}, as summarized in Fig. \ref{h_pn}.

On the other hand, the broadening of the spectra due to the static magnetic order was not observed for $0.6\le x \le 1$ of SrSc42622(As$_{1-x}$P$_x$), which corresponds to the AFM2 phase that reemerged at $0.4\le x' \le 0.7$ of LaFe(As$_{1-x'}$P$_{x'}$)O, in spite of $h_{\rm Pn}$ (1.2 \AA$<h_{\rm Pn}<$1.25 \AA) being similar.
We note that the slight difference in the two compounds occurs in the Fe-Fe bond length $d_{\rm Fe-Fe}$ contrary to the case of the AFM1 order phase, which is robust against the variation of $d_{\rm Fe-Fe}$. 
Based on the band calculation, it has been suggested that the AFM2 order of LaFe(As$_{1-x'}$P$_{x'}$)O is derived from the good Fermi surface (FS) nesting at bands mainly composed of $d_{xz}$/$d_{yz}$ orbitals, which accidentally improves at the intermediate $x'\sim0.6$ of LaFe(As$_{1-x'}$P$_{x'}$)O~\cite{Usui_SR}.  
Hence, the absence of the AFM2 phase in SrSc42622(As$_{1-x}$P$_x$) can be attributed to the deformation of the well-nested FSs by the slight difference in $d_{\rm Fe-Fe}$ even if the $h_{pn}$ is comparable. 

In contrast, it is noteworthy that the AFM1 phase seems to be insensitive particularly to $d_{\rm Fe-Fe}$, suggesting that the AFM1 and AFM2 phases have different origins. 
That is, the robustness of the AFM1 phase against the deformation of local lattice parameters such as $d_{\rm Fe-Fe}$ cannot be attributed only to the nesting of FSs, which is worse than that in the AFM2 phase, even though the FSs are composed of FSs of which the size of the hole and electron pockets originating from the $d_{xy}$ and $d_{xz\!/yz}$ orbitals are similar\cite{Mazin,Kuroki}. 
The band calculation over various Fe-pnictides suggested that the electron correlation effect on $3d$ orbitals becomes more significant when $h_{pn}$ is higher\cite{Miyake_U,Misawa}.
In this context, the robustness of the AFM1 order against the variation of local lattice parameters suggests that, rather than relying only on the nesting of FSs, the origin of the AFM1 phase largely relies on the electron correlation effect.

\begin{figure}[tbp]
\centering
\includegraphics[width=8cm]{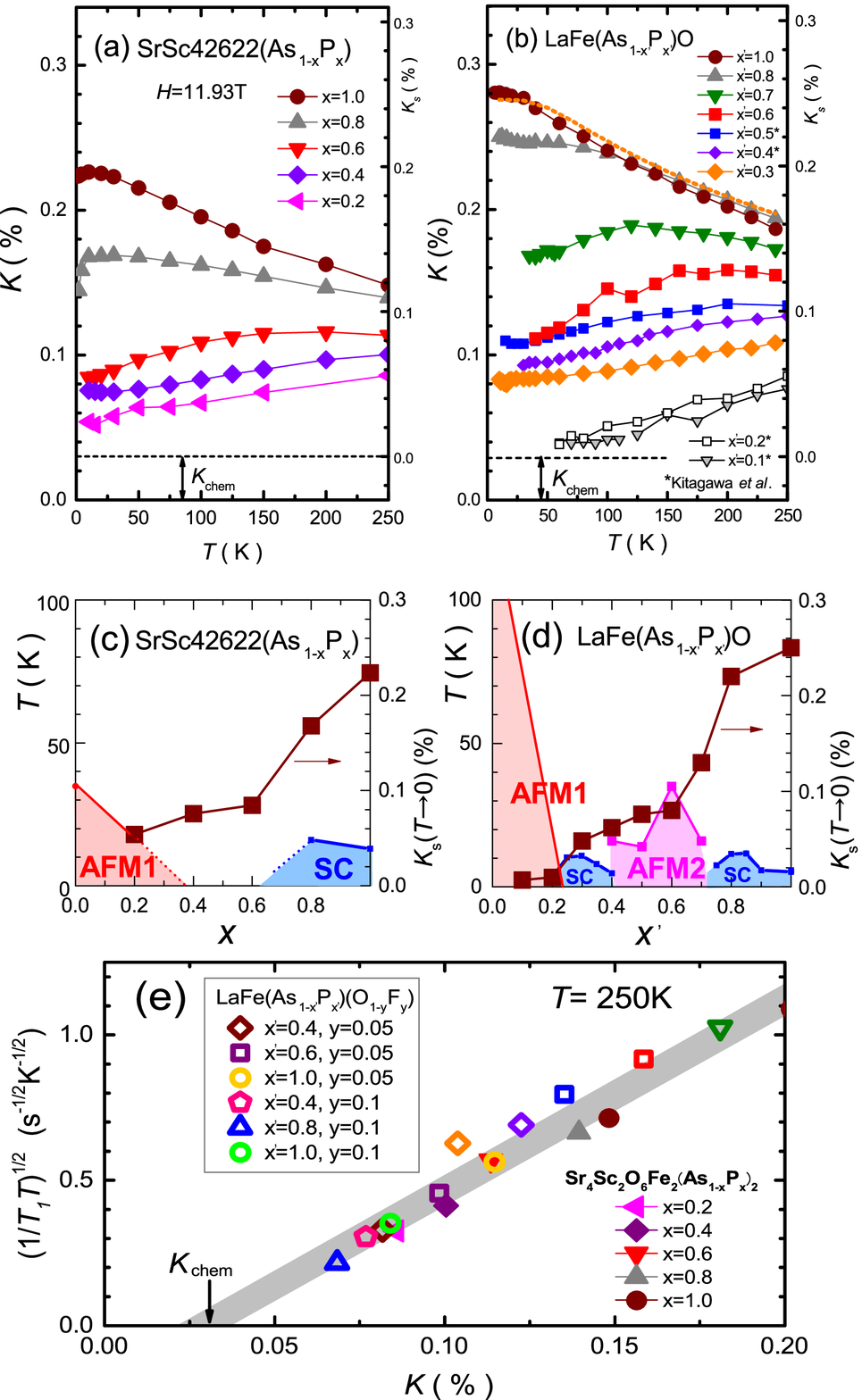}
\caption[]{(Color online) 
$T$ dependences of $K$($T$) probed by $^{31}$P-NMR for (a)  SrSc42622(As$_{1-x}$P$_{x}$) in comparison with that of (b) LaFe(As$_{1-x'}$P$_{x'}$)O~\cite{SKitagawa_2014,Lai_PRB,Miyasaka,Mukuda_jpsj2014}. 
The dependence of $K_s$($T\rightarrow$0) on $x$ estimated by extrapolation to $T$=0 in the paramagnetic state for (c) SrSc42622(As$_{1-x}$P$_{x}$) and (d) LaFe(As$_{1-x'}$P$_{x'}$)O. 
The large increase in $K_s$ upon cooling observed from $x$=0.6 to 1.0 for both compounds is due to the appearance of a sharp peak of DOS mainly from the $d_{3z^2-r^2}$ orbital\cite{Miyake,Mukuda_jpsj2014}.
(e) Plot of $(1/T_1T)^{1/2}$ vs. $K$ with an implicit parameter of $T$, together with data of LaFe(As$_{1-x'}$P$_{x'}$)(O$_{1-y}$F$_{y}$) in the paramagnetic states at $T\sim$250 K. 
The relation $(1/T_1T)^{1/2} \propto K_{\rm chem}+K_s(T)$ is linear with $K_{\rm chem}\sim$0.03\%.  
} 
\label{Knightshift}
\end{figure}

\subsubsection{Evolution of electronic states at 0 $\le x \le1$ evaluated from $^{31}$P-Knight shift } 

To focus on the difference and/or similarity between SrSc42622(As$_{1-x}$P$_x$) and LaFe(As$_{1-x'}$P$_{x'}$)O, we compare the phosphorous(P)-derived evolution of the electronic states through $^{31}$P-NMR probe. 
Figure \ref{Knightshift}(a) shows the $T$ dependence of Knight shift $K$ for each $x$. 
Here, $K$ comprises the $T$-dependent spin shift $K_s(T)$ and the $T$-independent chemical shift $K_{\rm chem}$.  
The former $K_s(T)$ is given by $K_s(T)\propto~^{31}\!A_{\rm hf}(0)\chi_0\propto~^{31}\!A_{\rm hf}(0)N(E_{\rm F})$, using the static spin susceptibility $\chi_0$, the density of states (DOS) $N(E_{\rm F})$ at the Fermi level $E_{\rm F}$, and the hyperfine coupling constant $^{31}A_{\rm hf}$(0) at ${q}$=0.  
In nonmagnetic compounds, $K_s(T)$ is anticipated to be proportional to $(1/T_1T)^{1/2}$, because Korringa's relation $(1/T_1T)\propto N(E_{\rm F})^2$ holds.  
The plot of $(1/T_1T)^{1/2}$ vs. $K$ in Fig.~\ref{Knightshift}(e) closely approximates the linear relation  $(1/T_1T)^{1/2} \propto K_{\rm chem}+K_s(T)$ with $K_{\rm chem}\sim$0.03 ($\pm$0.01)\%, which coincides with that evaluated in previous $^{31}$P-NMR studies for various Fe-pnictides\cite{Mukuda_PRB2014,Mukuda_jpsj2014,Miyamoto,Shiota}. 

As shown in Fig. \ref{Knightshift}(a) and \ref{Knightshift}(b), the $T$ dependence of $K_s(T)$ ( = $K(T) - K_{\rm chem}$) for each $x$ of SrSc42622(As$_{1-x}$P$_x$) is quite similar to that of  LaFe(As$_{1-x'}$P$_{x'}$)O\cite{SKitagawa_2014,Mukuda_jpsj2014}. 
The value of $K_s$($T\!\!\!\rightarrow$0) estimated from the extrapolation to $T\!\!\!\rightarrow$0 provides a direct measure of  $N(E_{\rm F})$ for a wide range of $x$, as shown in Fig.~\ref{Knightshift}(c), because $K_s$($T\!\!\!\rightarrow$0) is directly proportional to $\chi_0$ or $N(E_{\rm F})$. 
For $x\le0.6$ in both compounds, small values of $K_s$($T\!\!\!\rightarrow$0), i.e., $N(E_{\rm F})$, and a decrease in $K_s$($T$) upon cooling are characteristic, suggesting that the $E_{\rm F}$ is located on the tail of the large peak of the DOS beneath $E_{\rm F}$~\cite{Ikeda,Mukuda_jpsj2014}. 
At $x\ge 0.8$, as seen in the figure, $K_s$($T\!\!\!\rightarrow$0), i.e., $N(E_{\rm F})$, increases toward $x\rightarrow$1 owing to the appearance of the peak of DOS mainly arising from the $d_{3z^2-r^2}$-derived three-dimensional hole pocket around Z($\pi$,$\pi$,$\pi$)\cite{Miyake,Mukuda_jpsj2014}.
Note that the AFM2 phase in LaFe(As$_{1-x'}$P$_{x'}$)O does not exist in SrSc42622(As$_{1-x}$P$_x$) irrespective of the similar evolution in $N(E_{\rm F})$ as a function of $x$, suggesting the absence of any relation between the AFM2 phase and the quasiparticles on the $d_{3z^2-r^2}$ orbit. 
Instead, the presence of the cylindrical FSs originating from the nearly two-dimensional bands derived from the $d_{xz}$/$d_{yz}$ relation is suggested theoretically in both SrSc42622(P)\cite{Nakamura} and LaFePO\cite{Kuroki2}, although their fraction in the DOS is smaller than that of the $d_{3z^2-r^2}$ orbit. 
We note that, in the case of non-SC states of $A$Fe$_2$P$_2$ ($A$=Ba,Sr) being in normal metallic state, there is no features of highly cylindrical FSs,  electron correlations, and AFM spin fluctuations at all\cite{Kasahara,NakaiPRL,Kobayashi,KobayashiPRB2013,Miyamoto}. 
In the next section, we focus on the relation between AFM spin fluctuations and the onset of superconductivity in SrSc42622(As$_{1-x}$P$_x$).  

\begin{figure}[htbp]
\centering
\includegraphics[width=6.5cm]{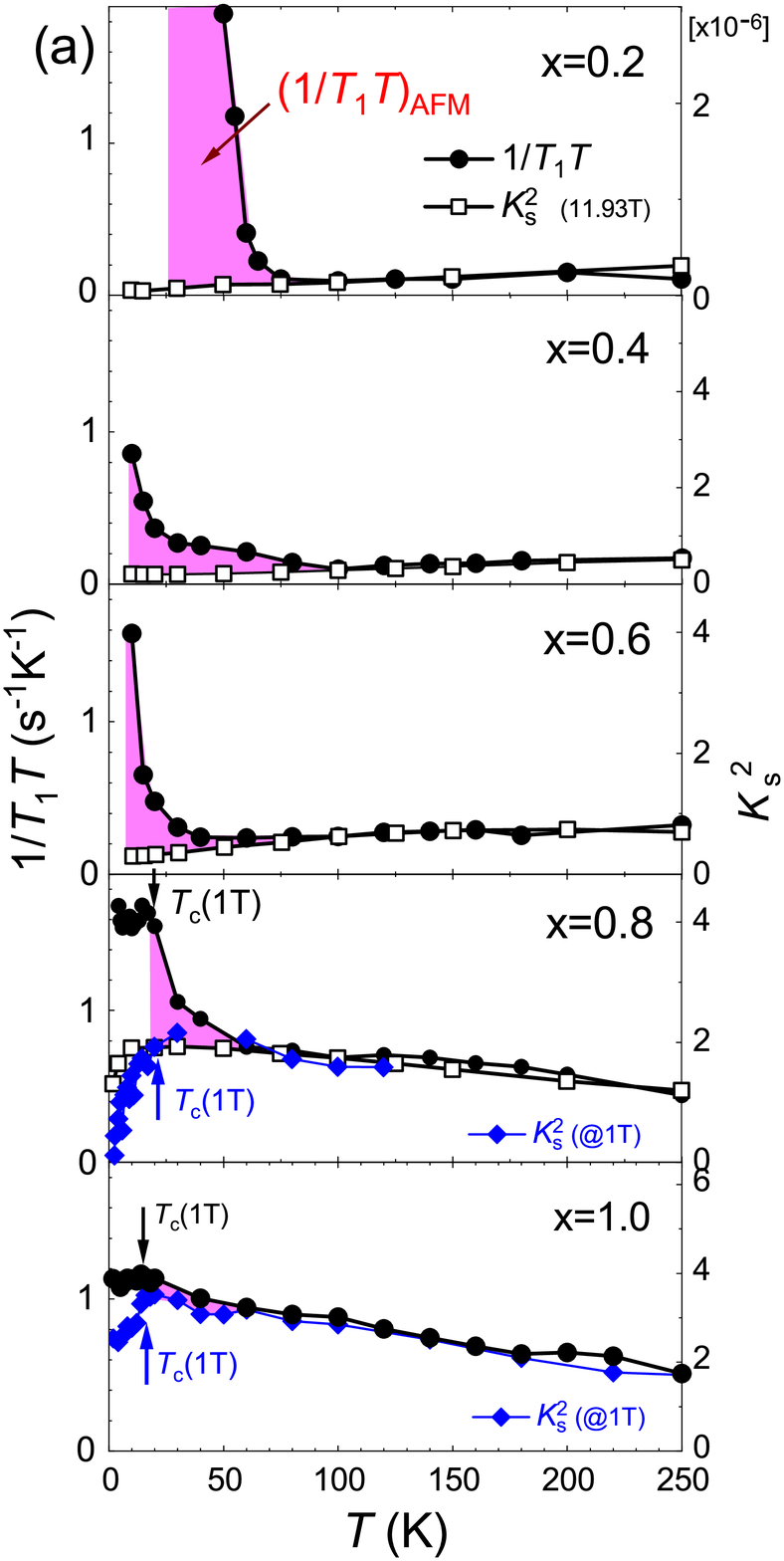}
\includegraphics[width=6.5cm]{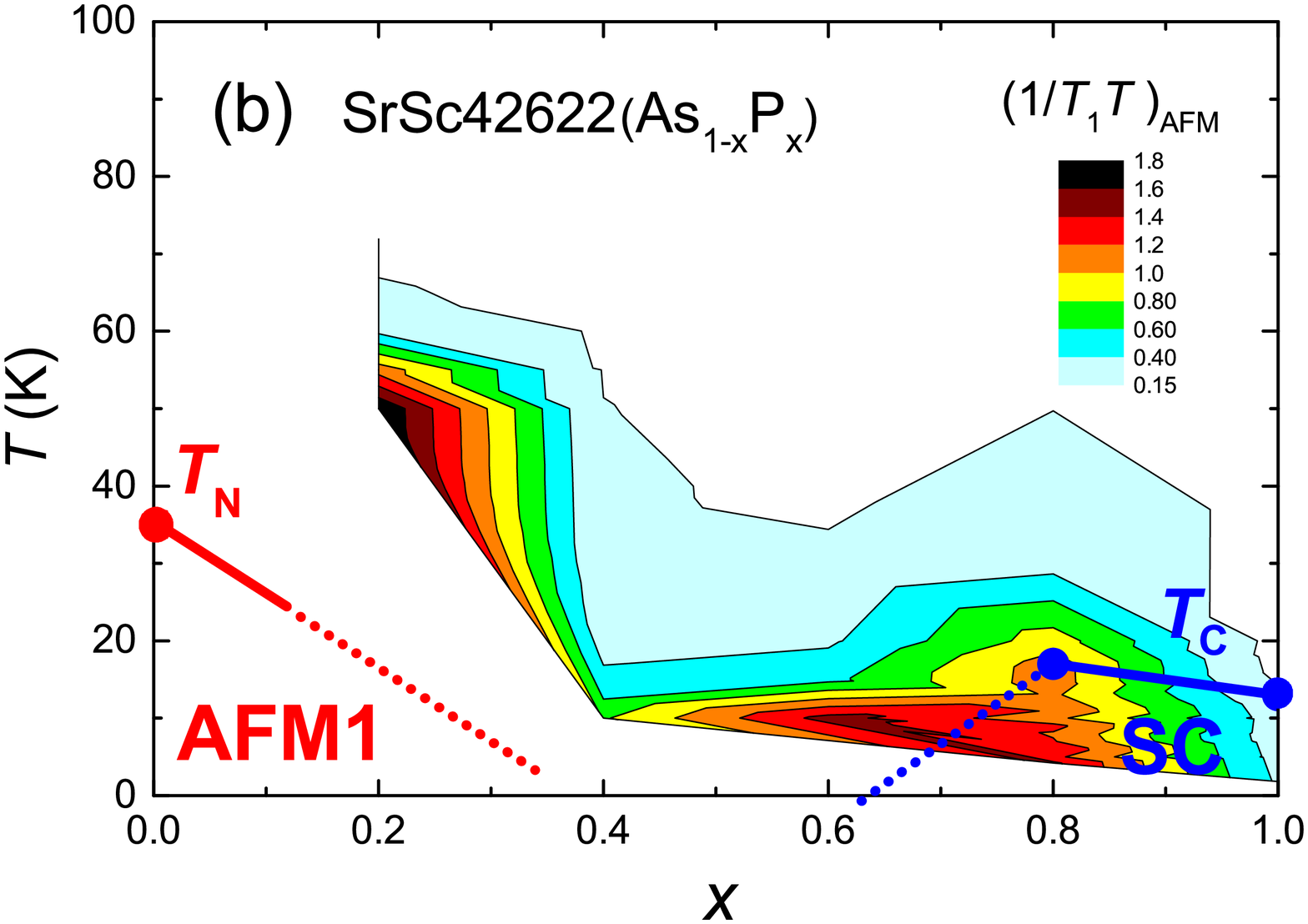}
\includegraphics[width=6cm]{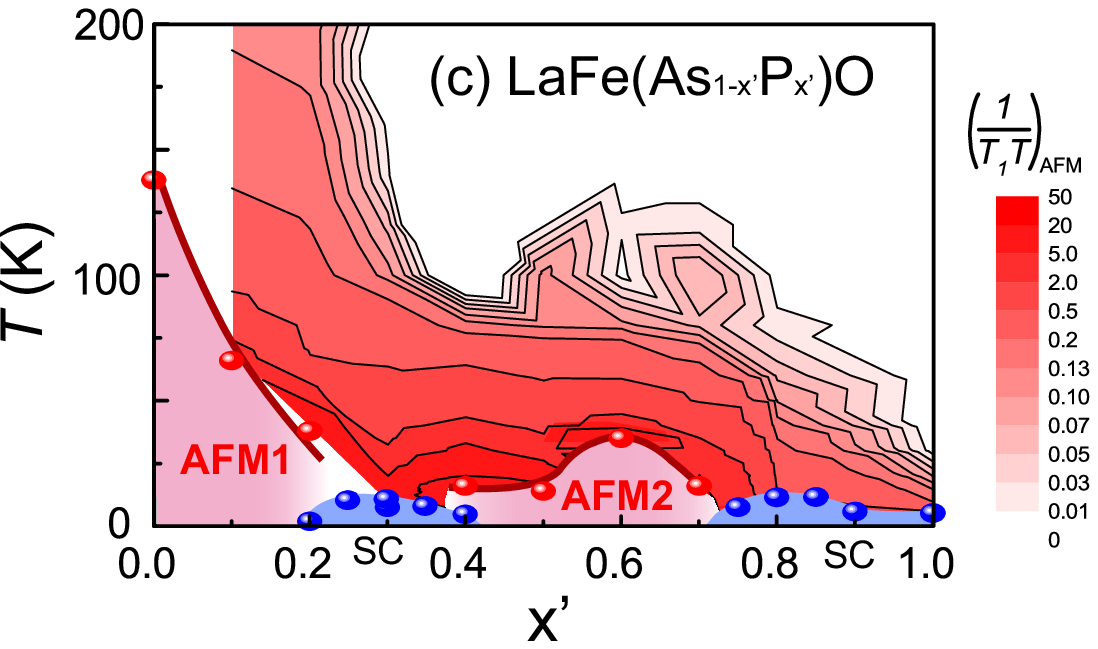}
\caption[]{(Color online) 
(a) $T$ dependences of $(1/T_1T)$ and $K_s^2(T)$ probed by $^{31}$P-NMR for  0.2$\le x \le$1.0.
The hatched area denoted as ($1/T_1T)_{\rm AFM}$ corresponds to the component of $1/T_1T$ due to the AFMSFs at a finite $Q$. 
The SC states at $x$=0.8 and 1.0 are measured at $B$=1 T, enabling us to detect the decrease in the Knight shift and the increase in the line width below $T_c$(1T).
(b) Contour plot of low-energy AFMSFs probed by ($1/T_1T)_{\rm AFM}$.
Evolution of AFMSFs upon cooling are observed significantly at $x$=0.8 with the highest $T_c$, which is discontinuous from that of the AFM1 phase. 
Here the broken line of the AFM1 phase is derived from the significant increase in FWHM of the spectra (See Fig. \ref{PNMR}(f)).  
(c) Contour plot of AFMSFs in LaFe(As$_{1-x'}$P$_{x'}$)O referred from a previous study.\cite{Shiota}
} 
\label{AFM-T1}
\end{figure}

\subsubsection{ AFM spin fluctuations and superconductivity} 

Figure \ref{AFM-T1}(a)  shows the  $T$ dependence of $(1/T_1T)$  and $K_s^2$  
probed by $^{31}$P-NMR for 0.2$\le x \le$1.0.
To deduce the development of AFM spin fluctuations (AFMSFs) following previous studies~\cite{Ning,Mukuda_PRB2014,Mukuda_jpsj2014,Miyamoto,Shiota}, we assume that $(1/T_1T)$ is decomposed as
\[
(1/T_1T)=(1/T_1T)_{\rm AFM}+(1/T_1T)_{0}.
\]
The first term $(1/T_1T)_{\rm AFM}$ represents the contribution of AFMSFs with the finite wave vectors $Q$, which is generally described as
\[
\left(\frac{1}{T_1T}\right)_{\rm \!\!AFM} \propto \lim_{\omega\rightarrow \omega_0\sim0}  |A_{\rm hf}(Q)|^2 \frac{\chi''(Q,\omega)}{\omega},
\]
where $A_{\rm hf}(Q)$ is the hyperfine-coupling constant at $q=Q$, $\chi(Q,\omega)$ is the dynamical spin susceptibility at $q=Q$ and energy $\omega$, and $\omega_0$ is the NMR frequency approximated as $\omega_0\sim 0$. 
The second term $(1/T_1T)_{0}$ represents the $q$-independent one in proportion to $N(E_{\rm F})^2$, which can be evaluated by $K_s^2(T)$($=$($K-K_{\rm chem}$)$^2$). 
The $T$ dependence of $(1/T_1T)$  above 100 K follows that of $K_s^2(T)$, indicating the dominant contribution of $(1/T_1T)_{0}$ to the observed $(1/T_1T)$.
Thus, the hatched area in Fig. \ref{AFM-T1}(a) corresponds to the evolution of $(1/T_1T)_{\rm AFM}$ that develops upon cooling below 100 K,  i.e., the evolution of AFMSFs at a finite ${Q}$ predominantly around $\sim$(0,$\pi$) and $\sim$($\pi$,0). 

The results are summarized in the contour plot of $(1/T_1T)_{\rm AFM}$ in  Fig. \ref{AFM-T1}(b).
At $x<$0.4, the AFMSFs evolve upon cooling toward possible magnetic order derived from the AFM1 phase, as suggested in Fig. \ref{PNMR}(g) by the increases in FWHMs. 
Although these are suppressed once at $x$=0.3-0.4, the AFMSFs are enhanced again significantly at $x$=0.6-1.0. 
Remarkably, the re-enhanced AFMSFs are not continuous from the AFM1 phase for $x<$0.4. 
This reminds us of the reemergence of AFMSFs and the {\it static} AFM2 order phase in the case of LaFe(As$_{1-x'}$P$_{x'}$)O, as referred in Fig. \ref{AFM-T1}(c)\cite{Shiota}.
We suggest that the AFMSFs at $x$=0.6-1.0 of SrSc42622(As$_{1-x}$P$_x$) can be attributed to the instability of the AFM2 order phase, because the local lattice parameters of their Fe$Pn$ layers are quite similar to that of the AFM2 phase for $x'$=0.4-0.7 of LaFe(As$_{1-x'}$P$_{x'}$)O\cite{SKitagawa_2014,Lai_PRB,Miyasaka,Mukuda_jpsj2014}.
In fact, it is theoretically suggested that the AFM2 phase is derived from the accidentally good nesting of the hole and electron FSs in the local lattice parameters of LaFe(As$_{0.4}$P$_{0.6}$)O\cite{Usui_SR}. 
Thus, these well-nested FSs in LaFe(As$_{0.4}$P$_{0.6}$)O may be missed by slightly longer  $d_{\rm Fe-Fe}$ in SrSc42622(As$_{1-x}$P$_x$), although the evolution of the microscopic electronic states evaluated by $K$ are quite similar to each other, as shown in Figs. \ref{Knightshift}. 
Despite the absence of the {\it  static} AFM2 order, the possible instability of the AFM2 phase induces the {\it  dynamical} low-energy AFM spin fluctuations that develop significantly at approximately $x\sim$0.8 of SrSc42622(As$_{1-x}$P$_x$).
We remark that such re-enhancement of the AFMSFs derived from the AFM2 order phase play a significant role in increasing the $T_c$ to 17 K at $x$=0.8$\sim$1, resulting in the highest $T_c$ among  phosphorous-rich Fe-based superconductors. 

In the SC state for both $x$=0.8 and 1.0, the bulk SC transitions are evidenced by decreases in $K_s(T)$ to below $T_c$(1T) under a low external field, $B$=1 T, as indicated by the arrows in Fig. \ref{AFM-T1}(a).
The $1/T_1T$ in the SC state shows the variation at $T_c$(1T) but its decrease is not  significant, which may be due to the large residual DOS at $E_F$ in the nodal SC gap\cite{Yates} resulting from the presence of the external field, since the field is not sufficiently low in comparison with the upper critical field $B_{c2}(0)$.  
Such large residual DOS in the nodal SC gaps are frequently reported in  $^{31}$P-NMR studies on phosphorous-rich members of the Fe-pnictides, such as LaFePO\cite{Nakai_LaFePO}, $M$Fe$_2$(As,P)$_2$\cite{NakaiPRL,Dulguun,Miyamoto}, and CaAl42622(P)\cite{KinouchiPRB}, which is more remarkable when three-dimensionality in the electronic structures is more significant, since the contribution from $d_{3z^2-r^2}$ orbit that has no correlation with SC is larger.
Further experiment  under zero field or low field, which is sufficiently lower than  $B_{c2}$, would be necessary.

\subsubsection{Origin of high $T_c$ state of FeP-based members } 

Here we compare the SC states and normal-state properties of the other FeP-based SC compounds, namely (i) [Ca$_4$Al$_2$O$_6$]Fe$_2$P$_2$(=CaAl42622(P))($T_c$=17 K)\cite{Shirage_AsP}, (ii) LaFePO ($T_c$=6 K)\cite{Kamihara2006}, (iii) LiFeP($T_c$= 5 K)\cite{Deng}, and (iv) $Ae$Fe$_2$P$_2$(non-SC). 
Their local lattice parameters are plotted in Fig. \ref{h_pn}.

(i) Among the FeP end members\cite{Shirage_AsP}, CaAl42622(P) exhibits the highest $T_c$ (=17 K) state, which appears in association with the low-energy AFMSFs in the vicinity of the AFM1 phase\cite{KinouchiPRB}. 
As indicated in Fig. \ref{h_pn}, this compound is characterized by FeP layers with relatively high $h_{Pn}$($\sim$1.3 \AA), which is close to the lower border of the AFM1 order. 
The nearly cylindrical FSs are ensured by the high two-dimensionality owing to the high $h_{Pn}$ and thick Perovskite blocking layer\cite{Usui42622}. 
Here no sign of the large DOS from the $d_{3z^2-r^2}$ orbit was observed\cite{KinouchiPRB}, contrary to SrSc42622(P) and LaFePO.  
The SC gap with nodes is clearly seen even at $B$=1 T.
(ii)  LaFePO with $T_c$=6 K appears with weak AFMSFs at the low energies derived from the AFM2 phase\cite{Mukuda_jpsj2014,Shiota}. 
In fact, their low-energy AFMSFs for $x'$=1 are more enhanced toward the AFM2 phase at $x'\le$0.7 in LaFe(As$_{1-x'}$P$_{x'}$)O, which brings about the doubly enhanced $T_c$(=12 K) for $x'$=0.8, as seen in Figs. \ref{Knightshift}(d) and \ref{AFM-T1}(c)\cite{Lai_PRB,Mukuda_jpsj2014,Miyasaka}.  
In addition to the weak AFMSFs at low energies observed in the previous NMR study\cite{Mukuda_jpsj2014}, the presence of the AFMSFs at high energies was reported for $x'\sim1$  from a neutron scattering experiment recently\cite{Ishikado}. 
Note that the AFMSFs probed by NMR correspond to the slope of the $\chi(Q,\omega)$ at the low-energy limit.
The cylindrical FSs are also ensured at some bands by the high two-dimensionality owing to the thick blocking layer\cite{Kuroki2}, although the DOS includes large components from the $d_{3z^2-r^2}$ orbit that may give rise to the large residual DOS in the nodal SC gaps analogous to the case of SrSc42622(P).
(iii) LiFeP with $T_c$= 5 K is characterized by higher $h_{Pn}$ and a thinner blocking layer, which retains the dominant two-dimensional FS properties, but causes one of the FSs to warp\cite{Shein,Ferber}.   
The occurrence of weak AFMSFs at low energies, which were reported in a previous NMR study\cite{Man}, may be attributed to the AFMSFs from the AFM1 phase because of the close proximity of $h_{Pn}$ to the border of the AFM1 phase as seen in Fig. \ref{h_pn}.
In this context, no SC was reported in (iv) $Ae$Fe$_2$P$_2$($Ae$=Ba,Sr), which may be attributed to that the lattice parameters in these compounds are far from both the AFM1 and AFM2 phases (See Fig. \ref{h_pn}). 
In fact, no indication of the presence of the AFMSFs, electron correlations, or two-dimensional features in their electronic structures was reported\cite{Kasahara,Kobayashi,NakaiPRL,Miyamoto}.
These experimental facts suggest that the presence of AFMSFs in the vicinity of either the AFM1 or AFM2 phases is an indispensable factor for the occurrence of SC in FeP-based compounds, although the multiband feature from the five $3d$-orbitals of Fe strongly diversifies the respective electronic states in detail. 

\begin{figure}[tbp]
\centering
\includegraphics[width=8cm]{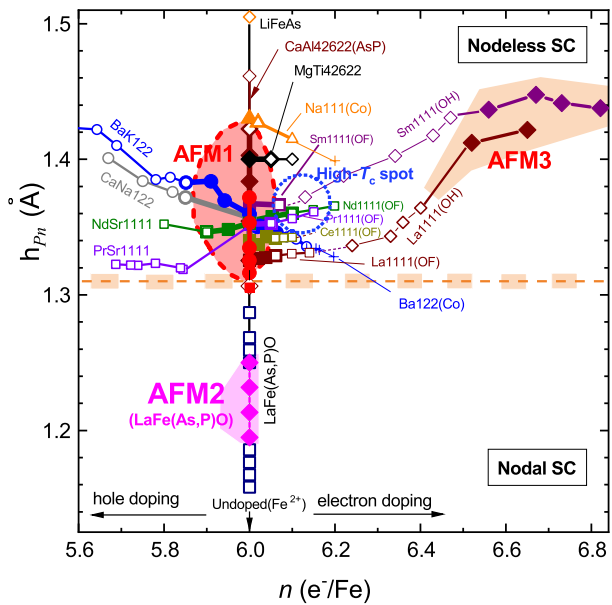}
\caption[]{(Color online)  
Ground state classified as functions of the doping levels of electrons/holes and the pnictogen height\cite{IshidaRev,Stewart,Scalapino,Hosono_review,Ren1,C.H.Lee,Mizuguchi,KinouchiPRB}. 
The undoped Fe$^{2+}$ state in Fig. \ref{h_pn} corresponds to a doping level of 6.0$e^-$ per Fe atom in this plot, and the electron- and hole-doping correspond to the region above and below 6.0$e^-$/Fe, respectively. 
The solid and open symbols represent the AFM order and the SC states, respectively. 
The AFM1 phase is widely observed in the parent Fe-pnictides for 1.3 \AA$<h_{Pn}<$1.42 \AA\  around 6.0$e^{-1}$/Fe, whereas the AFM2 and AFM3 phases are reported to occur only in LaFe(As,P)O\cite{SKitagawa_2014,Lai_PRB,Miyasaka,Mukuda_jpsj2014} and the heavily electron doped state $Ln$FeAs(O,H)\cite{Iimura_H,Hiraishi,IimuraSm}, respectively. 
Most of the nodeless and nodal SC phases in the Fe-pnictides are roughly classified by the border of the AFM1 and nodal SC phases around $h_{Pn}$=1.3-1.32 \AA\ \cite{Hashimoto}.
The region in which $T_c$ is the highest with $T_c>$50 K, is the region enclosed within a dashed circle, is the region in which the AFM1 phase is suppressed by electron-doping at $h_{\rm Pn}$=1.35-1.39 \AA\ in the middle of the AFM1 phase. 
Here we omit the data of SrSc42622(As$_{1-x}$P$_x$) on 6.0$e^{-1}$/Fe to avoid the overlap with the data of LaFe(As,P)O.  

}
\label{h_pn2}
\end{figure}

The highest $T_c$($\sim$55 K) in bulk Fe-pnictides is known to be realized by the optimization of both the local lattice parameters of the Fe$Pn$ layers and the electron-doping levels\cite{Ren1,C.H.Lee,Mizuguchi}.
Figure \ref{h_pn2} shows the AFM order and SC phases classified as functions of the doping levels of electron/holes and the $h_{\rm Pn}$. 
Here the doping level of 6.0$e^-$ per Fe atom corresponds to the undoped Fe$^{2+}$ state displayed in Fig.\ref{h_pn}, and sides to the right and left of 6.0$e^-$ correspond to the electron- and hole-doping regions, respectively.
The region in which $T_c$ is the highest with $T_c>$50 K, enclosed by the dashed circle in Fig.  \ref{h_pn2}, is the region in which the AFM1 phase is suppressed by electron-doping at $h_{\rm Pn}$=1.35-1.39 \AA\, which is in the middle of the AFM1 phase. 
The high $h_{\rm Pn}$ of the FeAs layers and suppression of the AFM1 phase by electron-doping can induce a fully gapped SC state on nearly cylindrical FSs \cite{Mazin,Kuroki,Kuroki2} and enhancement of the electron correlations and the AFMSFs\cite{Miyake_U,Misawa}, which may be more favorable for realizing higher $T_c$ states than in the FeP-based compounds. 

In contrast to the AFM1 phases widely observed in many parent Fe-pnictides, the universality of {\it static} reemergent AFM2 and AFM3 phases has not been established thus far, because these phases were reported only in LaFe(As$_{1-x'}$P$_{x'}$)O and the heavily electron-doped state $Ln$FeAs (O$_{1-y'}$H$_{y'}$)\cite{Iimura_H,Hiraishi,IimuraSm}, respectively. 
However, as for the relation of AFM1 and AFM2 phases, it was revealed that the $T_c$ can be enhanced not only by AFMSFs from either the AFM1 or AFM2 phases, but also by the superposition of multiple AFMSFs from both the AFM1 and AFM2 phases\cite{Shiota}. 
This is because the fluctuations from these two AFM phases are cooperative rather than competitive  \cite{Shiota}.
On the one hand, in the AFM3 phase of $Ln$FeAs(O$_{1-y'}$H$_{y'}$),  the hole FSs in association with the $d_{xz\!/yz}$  orbitals significantly shrinks owing to the heavy electron doping, whereas hole FSs relevant to the $d_{xy}$  orbit  and the large electron FSs still remain \cite{Iimura_H,Iimura_AF}. 
Theoretically it was suggested that the prioritized diagonal hopping on the $d_{xy}$ orbitals to re-enhance the other type of AFM order (AFM3) and AFMSFs in the high-$T_c$ state\cite{Suzuki_H}.
The highest-$T_c$ spot in Fe-pnictides with $T_c>$50 K appears between the AFM1 and AFM3 phases of $Ln$FeAs(O,F/H), and thus it is important to unravel the evolution of different types of AFMSFs from low energy to high energy systematically through the phases from AFM1 to AFM3 order \cite{Iimura_AF,Suzuki_H,Sakurai}. 
This would provide indications toward a universal understanding of the high-$T_c$ Fe-pnictides including large-electron FSs without a hole Fermi surface in monolayer FeSe compounds and intercalated FeSe-based compounds.\cite{Wang,He,Miyata,Zhao,Guo}

\section{Summary} 

A systematic NMR study on [Sr$_4$Sc$_2$O$_6$]Fe$_2$(As$_{1-x}$P$_x$)$_2$ revealed that the parent AFM1 phase at $x$=0 disappears in the range $x$=0.3-0.4, which corresponds to a pnictogen height from the Fe-plane of approximately $h_{pn}\sim1.3-1.32$\ \AA, which is nearly insensitive to the Fe-Fe bond length $d_{\rm Fe-Fe}$ for various {\it parent} Fe-pnictides.
By contrast, the {\it  static} AFM2 order reported to exist in  LaFe(As$_{0.4}$P$_{0.6}$)O does not appear at approximately $x\sim$0.8 of [Sr$_4$Sc$_2$O$_6$]Fe$_2$(As$_{1-x}$P$_x$)$_2$ although the local lattice parameters of the Fe$Pn$ layer are close to each other.
Despite the absence of the {\it  static} AFM2 phase, the {\it dynamical}  low-energy AFMSFs develop significantly at approximately $x\sim$0.8, and this development is discontinuous from that of the AFM1 phase. 
We remark that such re-enhancement of AFMSFs derived from the instability of the AFM2 order phase play a significant role in enhancing the $T_c\sim$17K at $x=$0.8$\sim$1. 
These results indicate that the onset of the {\it static} AFM2 order is quite sensitive to the local lattice parameters of the Fe$Pn$ layer, which is consistent with the anticipated fact that the AFM2 originates from the accidentally good nesting of FSs. 
This is in contrast with the case of the AFM1 phase, which is nearly insensitive to $d_{\rm Fe-Fe}$ when $h_{pn}\ge$1.3 \AA, at which the enhancement of electron correlations are suggested theoretically even when the nesting of FSs becomes weak. 
This suggests that the AFM1 and AFM2 phases have distinctly different origins.
However, the experimental facts suggest that the existence of AFMSFs in the vicinity of either the AFM1 or AFM2 phases is indispensable for the occurrence of SC in the FeP-based compounds. 
Although the multiband feature from the five $3d$-orbitals of Fe strongly and distinctly diversifies the respective electronic states, these findings provide insight into the complicated relationship between some segregated AFM order and the SC phases as a function of the local lattice parameters of the Fe$Pn$ layers. 
This advances the general understanding of the ground state of Fe-pnictides. 

\section*{Acknowledgements}

{\footnotesize 
We thank H. Usui and K. Kuroki for valuable discussion. 
This work was supported by the Izumi Science and Technology Foundation, Toyota Riken Scholar, and JSPS KAKENHI Grant Nos. 16H04013  and 18K18734. }



\begin{thebibliography}{99}
\bibitem{Tconset} Note that  $T_c$=13 K at $x$=1 in this work, which is determined by onset of diamagnetic response,  is slightly lower than $T_c$=17 K determined by the  onset of the resistivity drop\cite{Ogino}.
\end{thebibliography}
\end{document}